\documentclass[reprint,prb,aip]{revtex4-1}

\usepackage{graphicx}
\usepackage{bm}

\renewcommand{\vec}{\bm}

\begin{document}

\title{
Reversible tuning of omnidirectional band gaps in two-dimensional magnonic crystals by magnetic field and in-plane squeezing
}

\author{S.~Mamica}
	\email{mamica@amu.edu.pl}
	\affiliation{Faculty of Physics, Adam Mickiewicz University in Pozna\'n, ul.~Uniwersytetu Pozna\'nskiego 2, 61-614 Pozna\'n, Poland}

\author{M.~Krawczyk}
	\affiliation{Faculty of Physics, Adam Mickiewicz University in Pozna\'n, ul.~Uniwersytetu Pozna\'nskiego 2, 61-614 Pozna\'n, Poland}

\date{\today}

\begin{abstract}
By means of the plane wave method, we study nonuniform, i.e., mode- and $\vec{k}$-dependent, effects in the spin-wave spectrum of a two-dimensional bicomponent magnonic crystal. We use the crystal based on a hexagonal lattice squeezed in the direction of the external magnetic field wherein the squeezing applies to the lattice and the shape of inclusions. The squeezing changes both the demagnetizing field and the spatial confinement of the excitation, which may lead to the occurrence of an omnidirectional magnonic band gaps. In particular, we study the role played by propagational effects, which allows us to explain the $\vec{k}$-dependent softening of modes. The effects we found enabled us not only to design the width and position of magnonic band gaps, but also to plan their response to a change in the external magnetic field magnitude. This allows the reversible tuning of magnonic band gaps, and it shows that the studied structures are promising candidates for designing magnonic devices that are tunable during operation.
\end{abstract}

% ...

\maketitle

\section{Introduction}\label{sec_intro}

Magnonics is an important branch of research at present.\cite{Krawczyk_Grundler, Khitun_rev, Chumak_rev, Kruglyak_rev} The utilization of spin waves as information carriers is tempting but also challenging from different viewpoints, including basic research, application, and technology. In terms of applications, one of the important features of magnonic systems is the existence of the forbidden frequency range in the spin-wave spectrum, the so-called magnonic gap, which could be utilized as a stop band in spin-wave filters or transducers. From an application point of view, the crucial feature of the gap is its tunability. The occurrence of a band gap in the energy spectrum is the fundamental characteristic of periodic structures, including magnonic crystals (MCs).\cite{Vasseur_PWM, Nikitov_MCs} There are many opportunities to design magnonic band gaps in MCs by choosing their magnetic materials the MC consists of, as well as their structural parameters, such as dimensionality, periodic lattice type, or the shape of the inclusions.\cite{Krawczyk_SW, Vasiliev_SW, Wang_SW, Serga_SW, Lenk_SW, Tkachenko_SW, Klos_gap, Mamica_2D, Mandal_SW, Rychly_LTP, Choudhury2016, Rychly_SW, Goto_2019} However, the omnidirectional (complete) magnonic band gap opening in thin-film MCs requires rather high magnetic fields in the out-of-plane magnetized MCs \cite{Schwarze} or very large magnetization contrasts (Fe/Ni composites).\cite{Klos_gap} In in-plane magnetized permalloy or cobalt/permalloy MCs, only partial (directional) magnonic band gaps have been reported so far.\cite{Wang_SW, Kostylev, Tacchi_2DMC, Mruczkiewicz2013, Malago2015, Di2015, Gallardo2018, Gallardo2019}

The adjustment of material and structural parameters allows us to tailor the spin-wave spectrum, however the properties of the MC are fixed and a challenging task is to provide a tuning {\it in operando}. There are few approaches that address this problem. One of these approaches is called electrically controlled dynamic MCs.\cite{Chumak_DMC} The idea is to use the electric current flowing through an array of parallel wires to create a periodic external magnetic field in the vicinity of the uniform magnetic thin film. The field causes spin waves to propagate in a periodic potential the amplitude of which is tunable by the current. The width of the reported band gaps is in the range of a few tens of MHz, and the possibility of miniaturization is very limited. An all-magnetic method for the operational design of magnonic band gaps is to involve reprogrammable MCs in the form of a periodic array of magnetic stripes.\cite{Topp_gap, Tacchi_gap} The system can be reversibly switched between a ferro- and an antiferromagnetic configuration of the magnetization of particular stripes. The reported band gaps are up to c.a. 1~GHz wide, but the switching of these gaps is related to the magnetization reversal caused by the change in the external field direction. This results in hysteresis, large magnetic fields, and rather long times for the magnetization switching. Directional band gaps that are tunable by an external magnetic field without magnetization reversal were also reported in Fe/YIG composites, however the change in the gap width by 30\% requires a magnetic field up to 1.0~T.\cite{Ma2011}

In our recent paper,\cite{Mamica2019} we demonstrated another way to open the complete magnonic band gaps by squeezing the bicomponent MC structure. We found that in such structures the gap width can change significantly as a result of the small change in the external magnetic field magnitude (50--200 mT) without magnetization reversal. The mechanism involved is nonuniform mode softening.

In different magnetic systems at low magnetic field, a change in the field magnitude leads to a nonuniform frequency shift of the spin-wave spectrum. This has already been observed in one-dimensional MCs,\cite{Topp_gap, Tacchi_gap, Langer2017, Langer2019} in two-dimensional (2D) lattices of interacting magnetic dots,\cite{Montoncello_soft} and in anti-dot lattices.\cite{Zivieri_soft} The nonuniformity appears in two different effects. The first effect is a mode-dependent frequency shift and the second is a $\vec{k}$-dependent shift within the single band. The latter indicates a dependence on the direction and length of the wave vector $\vec{k}$. Such propagation effects can lead to several interesting features in planar magnetic systems, e.g., the virtual separation of a thin film into subsystems and the collapse of a band,\cite{HP1998, Mamica1998, Mamica2015} the mirage effect,\cite{Gruszecki2018} spin-wave lensing, and flow control.\cite{Whitehead2018, Gieniusz2017, Loayza2018, Krivoruchko2018, Krivoruchko2019}

The aim of the current work is to explain the influence of squeezing and an external field change on mode softening in the context of propagation effects. We study the mechanisms of the $\vec{k}$-dependent frequency shift in the spin-wave spectrum, and we show these mechanisms to be useful for the design and tuning of omnidirectional magnonic band gaps. We show that taking into account Gilbert damping does not change our main findings regarding both the existence of band gaps and the possibility of their tuning {\it in opperando}.

The structure is based on a squeezed hexagonal lattice wherein the squeezing affects both the lattice and the shape of the inclusions. We use the plane wave method (PWM) to calculate spin-wave spectra. In Sec. 2 we describe the basis of this approach and the model of the considered MCs. In Sec. 3 we examine the effect of squeezing on the spin-wave spectrum. We analyze the influence of the demagnetizing field, the spatial confinement of the spin-wave profile, and its concentration in cobalt/permalloy. In Sec. 4 we study propagational effects in the spin-wave spectrum. In particular, we explain the $\vec{k}$-dependent softening of modes. Finally, in Sec. 5, we provide some examples of 2DMCs that can be fabricated using current technology -- the examples, for which the existence and the behavior of omnidirectional band gaps are tailored by the squeezing of the structure. Here, we demonstrate the possibility of reversibly tuning magnonic band gaps by changing the external field magnitude, keeping the sample saturated. The gap width changes gradually at the rate designed by the MC structure. The paper is completed with conclusions (Sec. 6).

\section{The Model}

\begin{figure}
\includegraphics{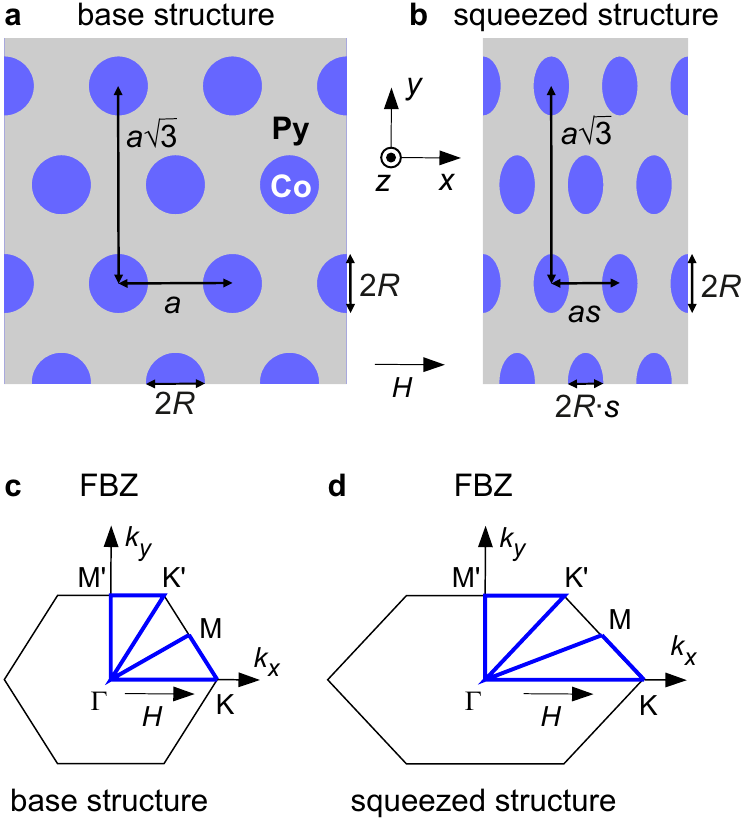}
\caption{2D MC based on the hexagonal lattice: cobalt rods (blue) in a permalloy matrix (gray). (a) The base structure: $a$ is the lattice constant, $R$ is the radius of Co rods. (b) The structure squeezed in the $x$ direction by the structure ratio $s$. (c, d) First Brillouin zone for the base and squeezed structures, respectively. Squeezing of the structure results in the proportional stretching of the FBZ. High symmetry paths are marked by blue lines.}
\label{Fig1}
\end{figure}

In this work, we study the spin-wave propagation in 2D MCs consisting of cylindrical cobalt inclusions (also called rods or dots) embedded in a thin-film permalloy matrix.\cite{Tacchi_2DMC, Duerr_JPD} The film thickness is 30 nm, and the material parameters used in this work are as follows: a saturation magnetization 1.39e6 A/m for Co, and 0.81e6 A/m for Py, an exchange stiffness constant 2.8e-11 J/m in Co, and 1.1e-11 J/m in Py, the Gilbert damping constant 0.002 for Co and 0.0063 for Py.\cite{Barati2014,Manago2015,Schoen216} The dots are arranged in sites of a 2D hexagonal lattice (Fig. \ref{Fig1}a). The lattice constant is $a = 600$ nm and the dot diameter is $R = 340$ nm. We will refer to this structure as the base structure. An external magnetic field $H$ is applied in the plane of the MC along the $x$ direction causing the demagnetizing field to rise at the interfaces between Co and Py. To manipulate this demagnetizing field the base structure is squeezed in the direction of the external field (Fig. \ref{Fig1}b). Although it is conceivable to squeeze the MC {\it in operando} \cite{Sadovnikov_straintronics} the aim of our work is to examine the influence of the demagnetizing field in the static case, i.e., the squeezed MC means the MC based on the squeezed structure (except for a short comment in Sec. 5). The squeezed structure will be described by the ratio of the new lattice constant in the $x$ direction to the original one which we will refer to as the structure ratio ($s$). Squeezing affects both the lattice and the shape of the rods. In Fig. \ref{Fig1}c and d the first Brillouin zone (FBZ) for the base and squeezed structure is provided, respectively, with blue lines used to mark the high-symmetry path. Please notice that squeezing of the structure leads to the elongation of the FBZ in the $k_{x}$ direction.

To describe the dynamics of the magnetization we use the classical continuous medium approach in which the equation of motion is the Landau-Lifshitz-Gilbert (LLG) equation:
\[
	\frac{ \partial \vec{M} }{ \partial t } = \gamma\mu_{0} \vec{M} \times \vec{H}_{\rm eff} + \frac{\alpha}{M_{S}} \vec{M} \times \frac{\partial\vec{M}}{\partial t} .
\]
Here $\vec{M} = \vec{M}\left(\vec{r},t\right)$ is the space- and time-dependent magnetization vector, $\gamma$ is the gyromagnetic ratio, and $\mu_{0}$ is the vacuum permeability. As in the case of free electrons, we assume $\left|\gamma\right|\mu_{0}=2.21 \cdot 10^{5}$ mA$^{-1}$s$^{-1}$. The second term describes the damping with the Gilbert damping parameter $\alpha$. In the effective magnetic field $\vec{H}_{\rm eff}$ we take into account three components: magnetostatic field, exchange field, and static external magnetic field $\vec{H}$. Assuming $\vec{H}$ to be strong enough to saturate the magnetization of the MC we use the linear approximation, thus $\vec{M}=\left[M_{S},m_{y},m_{z}\right]$, where $M_{S}$ is the saturation magnetization and $m_{z}$ and $m_{y}$ are two dynamic components of the magnetization vector. All vectors are expressed in the Cartesian coordinate system with the $x$--axis oriented along the external field, the $y$--axis laying in the plane of the MC, and the $z$--axis pointing in the out-of-plane direction (see the inset in Fig. \ref{Fig1}).

To solve the LLG equation we use the PWM which is a popular approach in magnonics that has already been elucidated in several papers (see, e.g., Refs.  \onlinecite{Vasseur_PWM, Krawczyk_3D, Romero_PWM, Banerjee_PWM, Rychly_PWM, Chang_PWM}), therefore here we recall only its main steps. After linearization of the LLG equation, the method involves two transformations which require ideal periodicity and allow us to consider only the unit cell with the periodic boundary conditions. All material parameters, such as saturation magnetization $M_{S}$,  exchange stiffness constant $A$, exchange length $\lambda_{ex}$, or damping parameter $\alpha$ are periodic in the real space and thus are Fourier-expanded. Bloch's theorem applies to the dynamic functions, such as the dynamic demagnetizing field components and the dynamic components of the magnetization. The final set of equations is equivalent to the following eigenvalue problem: \cite{Tiwari2010}
\begin{equation}\label{eq_problem}
\Omega 
\left[\begin{array}{cc}
\hat{1} & -\hat{\alpha} \\
\hat{\alpha} & \hat{1}
\end{array}\right]
\left[\begin{array}{c}
m_{y} \\
m_{z} 
\end{array}\right]
=
\hat{M}
\left[\begin{array}{c}
m_{y} \\
m_{z} 
\end{array}\right],
\end{equation}
with the matrices:
\[
\hat{\alpha}=\left[\begin{array}{ccc}
\ddots &  & 0  \\
 & \alpha\left(\vec{G}_{ii}\right) &  \\
0 &  & \ddots
\end{array}\right]_{N \times N} ,
\]
and
\begin{equation}\label{eq_matrix}
\hat{M}=\left[\begin{array}{cc}
\hat{M}^{yy} & \hat{M}^{yz} \\
\hat{M}^{zy} & \hat{M}^{zz}
\end{array}\right]_{2N \times 2N} ,
\end{equation}
$N$ being the number of plane waves used in the Fourier and Bloch expansion. For thin film bicomponent MCs with the external magnetic field applied in the plane of the film the elements of the matrix (\ref{eq_matrix}) are as follow:\cite{Rychly_LTP, Sokolovskyy_PWM, Mamica2018}
\begin{eqnarray}
M^{zz}_{ij} & = & - M^{yy}_{ij} = i \frac{k_{y}+G_{j,y}}{\left|\vec{k}+\vec{G}_{j}\right|} S\left(\vec{k}+\vec{G}_{j},z\right) M_{S}\left(\vec{G}_{i}-\vec{G}_{j}\right), \nonumber \\
M^{yz}_{ij} & = &  \frac{\left(k_{y}+G_{j,y}\right)^2}{\left|\vec{k}+\vec{G}_{j}\right|^2} \left(1-C\left(\vec{k}+\vec{G}_{j},z\right)\right) M_{S}\left(\vec{G}_{i}-\vec{G}_{j}\right) \nonumber \\
		& + & M^{\Sigma}_{ij}, \nonumber \\
M^{zy}_{ij} & = & - C\left(\vec{k}+\vec{G}_{j},z\right) M_{S}\left(\vec{G}_{i}-\vec{G}_{j}\right) - M^{\Sigma}_{ij}, \nonumber 
\end{eqnarray}
where the following symbols are used:
\begin{eqnarray}
&& M^{\Sigma}_{ij} = H \delta_{ij} \nonumber \\ 
&& - \frac{\left(G_{i,x}-G_{j,x}\right)^2}{\left|\vec{G}_{i}-\vec{G}_{j}\right|^2} \left(1-C\left(\vec{G}_{i}-\vec{G}_{j},z\right)\right) M_{S}\left(\vec{G}_{i}-\vec{G}_{j}\right) \nonumber \\ 
&& + \sum_{l}{\left(\vec{k}+\vec{G}_{j}\right) \cdot \left(\vec{k}+\vec{G}_{l}\right) M_{S}\left(\vec{G}_{i}-\vec{G}_{l}\right) \lambda_{ex}^{2}\left(\vec{G}_{l}-\vec{G}_{j}\right)   }, \nonumber
\end{eqnarray}
\begin{eqnarray}
S \left(\vec{k},z\right) & = & \sinh(\left|\vec{k}\right|z) \exp(\left|\vec{k}\right|d/2), \nonumber \\
C \left(\vec{k},z\right) & = & \cosh(\left|\vec{k}\right|z) \exp(\left|\vec{k}\right|d/2). \nonumber
\end{eqnarray}
Here, $M_{S}\left(\vec{G}\right)$, $\lambda_{ex}\left(\vec{G}\right)$, and $\alpha\left(\vec{G}\right)$ are Fourier expansions of the respective material parameters. The subscripts $i$, $j$, and $l$ are integer numbers from 1 to $N$, vectors $\vec{G}$ are reciprocal lattice vectors, and $\vec{k}$ is the Bloch wave vector of the SW. The exchange length is defined as $\lambda_{ex}=\sqrt{2A/\mu_{0}M_{S}^{2}}$.\cite{Krawczyk_PWM} The above formulas are valid for the excitations uniform in the $z$ direction, which is a reasonable assumption for a film as thin as $d=30$~nm, as it is in our case.

Fourier expansions of each material parameter $Y$ 
are  
$
Y\left(\vec{G}\right) = \left(Y_{\rm{Co}}-Y_{\rm{Py}}\right) f \frac{2J_{1}\left(GR\right)}{GR}
$
for $\vec{G} \neq 0$ and 
$
Y(0) = \left(Y_{\rm{Co}}-Y_{\rm{Py}}\right) f + Y_{\rm{Py}}
$
for $\vec{G} = 0$, where $f$ is a filling fraction, which for squeezed hexagonal MCs studied in this paper is given by $f=\frac{\pi R^{2} }{ a^{2} \sin(\pi / 3) }$, and $J_{1}$ is a Bessel function of the first kind. The symbol $GR = \sqrt{ (G_{x}R_{x})^{2} + (G_{y}R_{y})^{2} }$, with $R_{x}$ and $R_{y}$ being semi-axes of the ellipse in the $x$ and $y$ direction, respectively.

Providing a numerical solution of the eigenproblem (\ref{eq_problem}) one can calculate reduced frequencies (eigenvalues) $\Omega = i \omega / \left|\gamma\right|\mu_{0}$ and eigenvectors $\vec{m}_{\vec{k}}\left(\vec{G}\right)$. The first provides information about spin-wave frequency (the real part of $\omega$) and life time (the imaginary part of $\omega$), which is approximately equal to the full width at half-maximum of the spectral line.\cite{Zakeri2007} The second are coefficients of the Bloch expansion of the dynamic magnetization component:
\[
	\vec{m}\left(\vec{r}\right) = \sum_{\vec{G}} \vec{m}_{\vec{k}}\left(\vec{G}\right) \exp\left(i\left(\vec{k}+\vec{G}\right)\cdotp\vec{r}\right).
\]
From this equation, one can calculate the spatial distribution of the dynamic magnetization for a given mode and $\vec{k}$, i.e., the SW profile. Usually, distributions of the $z$ and $y$ components obtained for the same mode are similar, so it is sufficient to provide just one component (in-plane or out-of-plane) to explain the character of the mode. In this work, we use $N = 271$ plane waves for Bloch and Fourier expansions, which is large enough to ensure the satisfactory convergence of the results.

Following Ref. \onlinecite{Mamica_mFT} we introduce a concentration factor which for rods reads:
\begin{equation}\label{eq_cf}
 cf_{A} = \frac{ \tilde{m}_{A} }{ \tilde{m}_{A} + \tilde{m}_{B} } .
\end{equation}
In the case of 2D MCs the mean value of the squared amplitude of the dynamic magnetization in the area $S_{X}$
 is $ \tilde{m}_{X} = \frac{1}{S_{X}} \int_{S_{X}} |\vec{m}|^{2} dS $ where for rods $X = A$ and for the matrix $X = B$. The quantity given by Eq. (\ref{eq_cf}) allows us to determine the dominant excitation distribution in different materials of any particular spin-wave mode, e.g., the in-rods concentration factor above 0.5 means that the concentration of dynamic magnetization is higher in Co rods than in the Py matrix.

\section{Consequences of the squeezing}

\begin{figure*}
\includegraphics{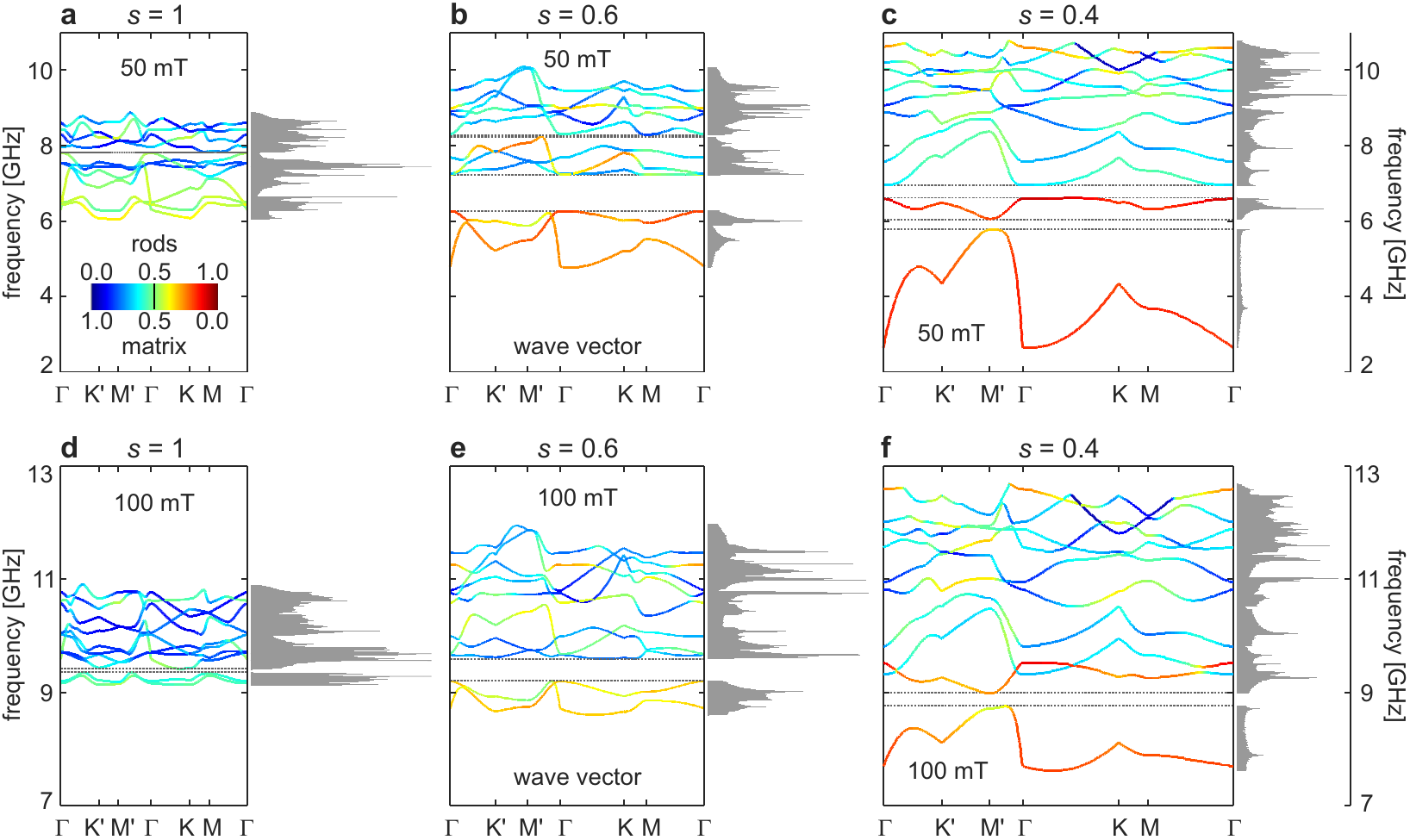}
\caption{Ten lowest modes in spin-wave spectra of Co/Py 2D MCs along the high symmetry path in the FBZ (Fig. \ref{Fig1}c and d) for two external magnetic field magnitudes: (a-c) 50 mT and (d-f) 100 mT, and for three squeezes: (a, d) the base structure (the structure ratio $s = 1$), (b, e) $s = 0.6$, and (c, f) $s = 0.4$. Line colors depict the concentration factor calculated from Eq. (\ref{eq_cf}) according to the color scale shown in the inset of (a). Dotted horizontal lines represent the upper and lower band edges of complete band gaps. All graphs are accompanied by the density of states calculated over the whole FBZ and plotted to the left all on the same scale.}
\label{Fig2}
\end{figure*}

\begin{figure*}
\includegraphics{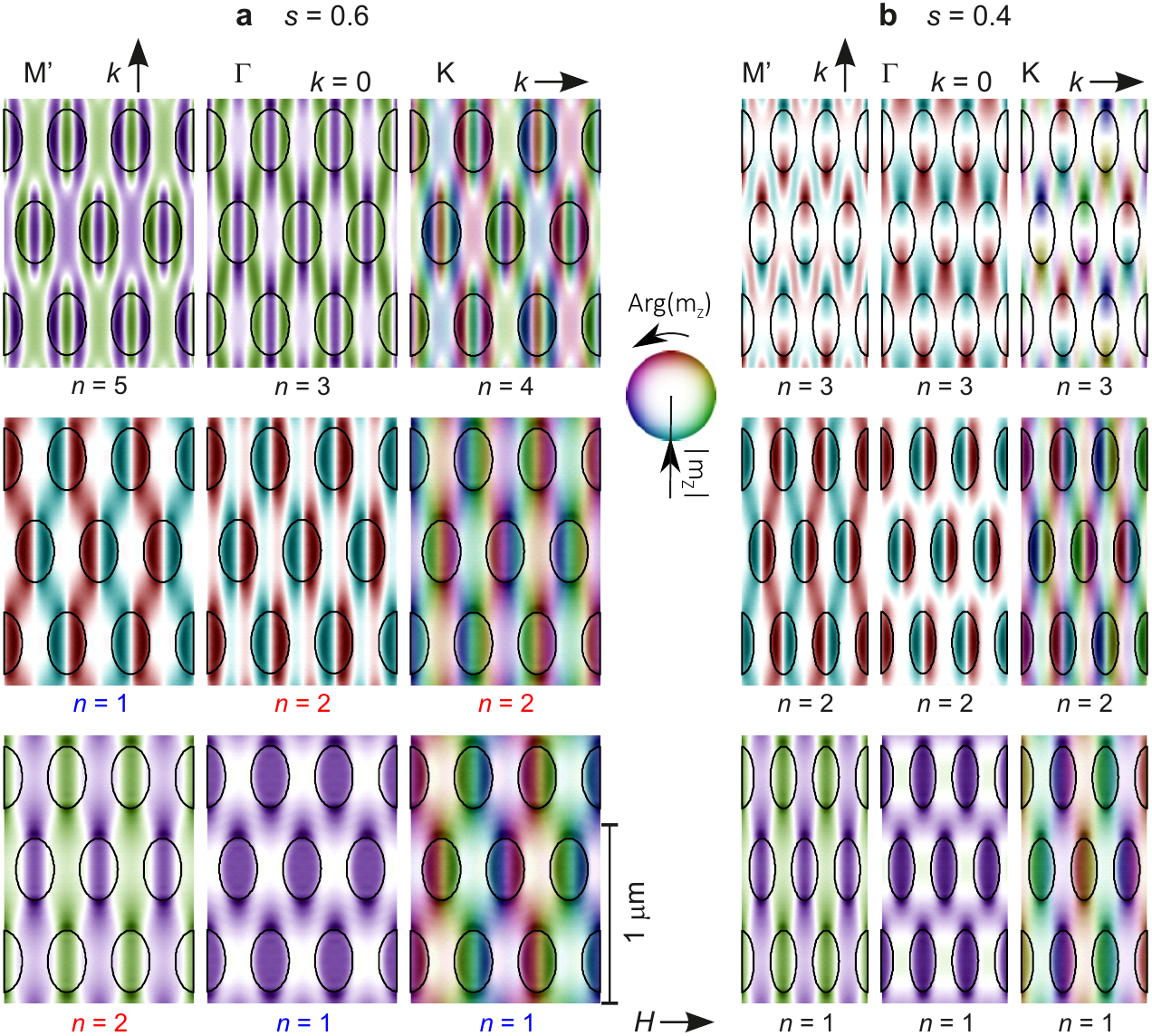}
\caption{Spin-wave profiles of three modes with the lowest frequency at the FBZ center for the structure ratio (a) $s = 0.6$ and (b) $s = 0.4$ at the external field 50 mT calculated at three high symmetry points in FBZ: M' (left column), $\Gamma$ (central column), and K (right column). Profiles in the same row correspond to each other and they are not always ordered according to their frequency in the spectrum (see mode number $n$). Ellipses mark Co rods borders. Colors represent argument (phase) and their intensity indicates the modulus of the dynamic magnetization, as is shown in the inset.}
\label{Fig3}
\end{figure*}

In Fig.~\ref{Fig2} we show representative spin--wave spectra calculated over the FBZ along the high symmetry path (Fig. \ref{Fig1}c and d; please notice that the squeezing of the structure leads to the elongation of the FBZ in the $k_{x}$ direction). Since we are interested in the widest band gaps which open at the bottom part of the spectrum, only the ten lowest modes are shown. Dispersion curves are colored to depict the concentration factor in the rods/matrix (see Eq. (\ref{eq_cf})) of each mode vs. the wave vector (the color scale is given in the inset in Fig. \ref{Fig2}a). Complete magnonic band gaps are marked by dotted horizontal lines drawn for their bounding frequencies. The spectra are arranged in two rows for two magnitudes of the external field: 50 mT (upper row) and 100 mT (lower row). Each row contains spectra for three structure ratios: 1.0 (base structure), 0.6, and 0.4. There is a density of states (DoS) plotted to the left of each spectrum, calculated over the entire FBZ as a direct summation of states in each small range of frequencies (all plots are in the same scale).

For the base structure at 50 mT (Fig. \ref{Fig2}a), the bunch of ten lowest modes fits in the range between 6 and 9 GHz. The concentration factor slightly depends on the wave vector. The spin-wave spectrum exhibits a very narrow band gap of width 22 MHz just below 8 GHz. The gap occurs between the 6th and 7th mode (we will refer to as the 6th gap, because it appears above 6th mode). For the squeezed structure (Fig. \ref{Fig2}b) this gap does not occur but two other band gaps exist instead: the 961 MHz wide 2nd gap and the 5th gap, which is as narrow as 53 MHz. The spectrum, except for the two lowest-frequency modes, is shifted up and now the top frequency is c.a. 10 GHz. The two lowest modes behave completely differently: their frequency range widens and shifts down. Additionally, there is a change in the concentration factor that is notably visible for these two modes (for the lowest mode in the FBZ center $cf_{A}$ changes from 0.53 to 0.73) and the variation of the concentration factor with the wave vector increases. For a more squeezed structure ($s = 0.4$, Fig. \ref{Fig2}c) the concentration factor is much more diverse, which results in the $\vec{k}$-dependent shift of frequencies and finally in the closing of the 5th gap. The in-rod concentration of the two lowest modes continues to grow, however, only the lowest one moves down in the spectrum while the second mode shifts up. As a consequence, the 2nd gap narrows down to 339 MHz, but the 1st gap appears around 6 GHz and its width is 273 MHz. The rest of the spectrum again is shifted up. The similar behavior of the spin-wave spectrum represents an external field of 100 mT (compare Figs. \ref{Fig2}d--f).

Calculation of the DoS gives the same range of forbidden frequencies as the spectra over the high symmetry path (Fig. \ref{Fig2}). Additionally, for isolated bands, its reveals some information about group velocity -- a higher DoS means a flattening of the frequency branch and a lower group velocity. An interesting feature is that some of the DoS maxima are located inside the frequency band instead of its border. The different behavior of the two lowest modes upon squeezing is clearly reflected in their DoS.

In our previous paper devoted to squeezed 2D MCs, we noticed the influence of the growing demagnetizing field upon the squeezing of the structure.\cite{Mamica2019} Due to the higher saturation magnetization of Co, the demagnetizing field in rods is negative while in the matrix it is positive. The squeezing of the structure along the external field direction causes both absolute values to grow, which leads to a stronger reduction in the internal field (defined as the sum of the external magnetic field and the demagnetizing field \cite{Demokritov_book}) in rods, and it makes spin waves easier to excite there. Thus, the intensification of the demagnetizing field influences rod-concentrated modes in two ways: by a direct lowering of the frequency due to the reduction in the internal field, and by the growth of the concentration factor, which enhances the first effect. Finally, for $s = 0.6$ we can distinguish two modes separated from the rest of the spectrum, causing an almost 1 GHz wide gap to open (Fig. \ref{Fig2}b). Both modes are highly concentrated in Co rods. For stronger squeezing the lowest mode continues to soften while the second mode stops softening and shifts up (Fig. \ref{Fig2}c). In Ref. \onlinecite{Mamica2019} we explain this behavior by examining the spin-wave profiles for the modes in question: the frequency of higher order modes, which have more nodal lines in their profiles, shifts up while the structure is squeezed due to the quantization in a smaller area (an effect similar to the wave confined in the potential well). Thus, the final frequency shift is the result of the competition between two opposite effects, namely `softening' of cobalt rods and growing of the spatial confinement. These findings helped us to formulate a rough explanation of the mode-dependent shift of frequencies in the spin-wave spectrum visible in Fig. \ref{Fig2}.

We will now explore the problem in greater detail. In Fig. \ref{Fig3} we plot spin-wave profiles for two squeezed structures: $s = 0.6$ (panel a) and $s = 0.4$ (panel b). In both panels, there are three columns corresponding to three high symmetry points in the FBZ (see Fig. \ref{Fig1}c). The left column is for point M', i.e., the FBZ border for the propagation in the $y$ direction (bottom-up in the pictures), the central column is for the FBZ center ($\Gamma$, $k = 0$), and the right column is for point K lying at the FBZ border for  propagation in the $x$ direction--the direction of the external field $H$ (left-right in the picture). The profiles of the same mode are organized in rows. The central-column profiles are for modes with successive frequencies in the FBZ center ($n =$ 1, 2, 3). Due to the mode crossing at points M' and K, the mode order can be different, as in the case of Fig. \ref{Fig3}a. For $s = 0.4$, the modes are well separated and their order does not change.

For both structure ratios, two lowest modes in point $\Gamma$ (central column) are strongly concentrated in Co rods, and their profiles are similar. The lowest one is a fundamental mode--a counterpart of the uniform excitation having magnetization precessing all in phase (which is not necessarily the lowest mode in the spectrum).\cite{MamicaDot2012, MamicaDot2013, MamicaDot2014} The second mode exhibits two nodal lines and a phase change $2\pi$ between neighboring rows of Co in the $x$ direction (going from left to right crosses nodal lines), i.e., the direction of the squeezing. However, within a single rod the phase changes by $\pi$, leading to one nodal line (while there are two in the matrix). This supports our previous conclusion concerning the mode-dependent shift of the frequency upon squeezing: this phase change makes the frequency more sensitive to  spatial confinement, and thus it forces it to increase for stronger squeezing. The third mode (in $\Gamma$) for $s = 0.6$ ($n = 3$ in Fig. \ref{Fig3}a) can be treated as a next-order mode inside rods. The profile exhibits one full period of oscillation between adjacent lines of rods in the $x$ direction: the phase changes by $2\pi$ with two nodal lines. This is similar to the profile of the mode $n = 2$ but now anti-nodal lines are shifted close to the position of nodal lines for $n = 2$, and two nodal lines occur resulting in the phase changes by $2\pi$ inside a single rod. Thus, the mode is more sensitive to the spatial in-rod confinement which pushes up its frequency more than for $n = 1$ or 2. Further squeezing shifts this mode in the spectrum more than other modes causing the mode reordering and for $s = 0.4$ another mode is the third one in $\Gamma$ ($n = 3$, Fig. \ref{Fig3}b). This mode is concentrated mostly in the matrix. Its profile has one nodal line inside rods but in the $y$ direction (and another one in the matrix in the $x$ direction). So there is a full $2\pi$ phase change every two rows of rods in both directions.

\section{Propagational effects}

First, let us introduce some abbreviations: RD – the row of Co dots (rods), ND – neighboring Co dots, both can be counted in the $x$, $y$ or $o$ (oblique, $\pm 60^{\circ}$) direction. For instance, RDy means a row of rods along the $y$ direction (a vertical row in Fig. \ref{Fig3}) or NDo means two rods neighboring in an oblique direction.

In the case of propagating modes (columns M' and K in Fig. \ref{Fig3}), an additional phase change appears due to the non-zero wave vector. This change depends on the direction and length of the wave vector and interact with the phase distribution for $k = 0$, and it has a big impact on the spin-wave frequency. Both of the lowest modes ($n = 1$) presented in Figs. \ref{Fig3}a and b are all-in-phase in the FBZ center. The point M' stands for the propagation in the $y$ direction with the wave vector at the FBZ border, i.e., in the middle between two neighboring sites of the reciprocal lattice. So, there should be an additional $\pi$ phase change between two successive RDx's. Indeed, inside rods from neighboring RDx's there is a phase change by $\pi$ but due to the shift in the $x$ direction by half of the lattice constant the opposite phases are shifted also in the $x$ direction. (Because the mode is strongly concentrated in rods the profile in the matrix is less important.) The final effect is, that the profile consists of all-in-phase oscillations along RDy's with alternating phase in the $x$ direction (even if the propagation is in the $y$ direction). The alternation of the phase causes the frequency to increase and this effect depends on the squeezing (spatial confinement): the increase in the frequency between points $\Gamma$ and M' for $s = 0.6$ is c.a. 1 GHz while for $s = 0.4$ around 3 GHz (Figs. \ref{Fig2}b and c). At point K, one reaches the FBZ border for the wave propagation in the $x$ direction. In the reciprocal space this point is located at the distance of 2/3 between two neighboring sites. Thus the additional change of phase due to the non-zero wave vector is $2\pi$ every three RDy's. Combining this with the fundamental mode for $s = 0.6$ results in a continuous change in the phase seen in the bottom-right panel of Fig. \ref{Fig3}a. The phase is changing both in the rods and the matrix, albeit more rapidly in the rods. This causes the frequency to increase but to a lesser extent than at point M' (Fig. \ref{Fig2}b). For a more squeezed structure ($s = 0.4$) the phase inside the rods changes little, therefore a rapid change in the phase appears between neighboring RDy's (the bottom-right panel in Fig. \ref{Fig3}b) and the frequency increases much more than in the previous case (Fig. \ref{Fig2}c).

The second mode at point $\Gamma$ consists of excitations strongly concentrated inside the rods and in close vicinity for both squeezed structures ($n = 2$, Figs. \ref{Fig3}a and b). In each rod there is a nodal line in the middle in the $x$ direction (the $x$ direction crosses the nodal line). Since all rods have the same phase on the left and the opposite phase on the right, there is an extra nodal line in the matrix separating any pair of NDo's. At point M', additional phase change by $\pi$ (resulting from the propagation) flips excitations in neighboring RDx's. Again, due to the lateral shift of neighboring RDx's the flipped phase is shifted in both Cartesian directions. Therefore, the phase does not change between neighboring RDy's (only in the middle of the RDy) and the nodal line between NDo's is removed. Fewer nodal lines in the matrix result in the lowering of the frequency at point M' (Figs. \ref{Fig2}b and c). In combination with the rise of the frequency of the first mode this results in a mode order change for $s = 0.6$. (For $s = 0.4$ the separation of these modes is large enough to prevent a change in their order.) The propagation in the $x$ direction (point K in the FBZ) brings two features: a continuous change in the phase and the disappearance of nodal lines--the phase changes smoothly. These two features have the opposite effect on the frequency which results in its slight change on the path from $\Gamma$ to K in the FBZ.

The third mode for $s = 0.6$ joins the higher part of the spin-wave spectrum and due to its frequency change over the FBZ the mode ascends in the spectrum in both directions of propagation: at M' it becomes the fifth mode and at K the fourth one. The change in its profile is shown in the upper row of Fig. \ref{Fig3}a. The profile for point $\Gamma$ in the $x$ direction exhibits one full period of oscillation between adjacent RDy's: the phase changes by $2\pi$ with two nodal lines in the matrix but also inside of any rod. Because there is the same phase at both sides of every rod (green areas), the maxima of this phase form lines not only in rods but also in the matrix connecting NDy's. At point M' the additional phase change by $\pi$ in the $y$ direction causes a change in the phase around the rods from neighboring RDx's. As a result, there are no more anti-nodal lines connecting NDo's--the additional nodal line appears in the matrix which cause the frequency to increase. At point K, similarly to the second mode, nodal lines are removed and the phase changes smoothly. However, the phase change is more rapid, especially inside rods, which pushes the frequency up. The third mode for $s = 0.4$ is different (Fig. \ref{Fig3}b) which we explain in Sec. 3, but again, at point M' an additional nodal line appears in the matrix shifting the frequency up (Fig. \ref{Fig2}c). At point K the nodal line in the matrix is retained with an additional phase change in-between neighboring RDx's so the frequency increases.

Also, the in-rod concentration is nonuniform and anisotropic, i.e., it depends on both the length and the direction of the wave vector. This is related to the in-plane structure of the MC so for higher squeezing it should be stronger, resulting in a much more diverse concentration factor. Of course, the final result depends on the mode profile: any additional nodal line in the rods makes the excitation of the spin wave more difficult and pushes the profile out. The same applies to the matrix.

\section{Omnidirectional band gaps tailoring}

\begin{figure*}
\includegraphics{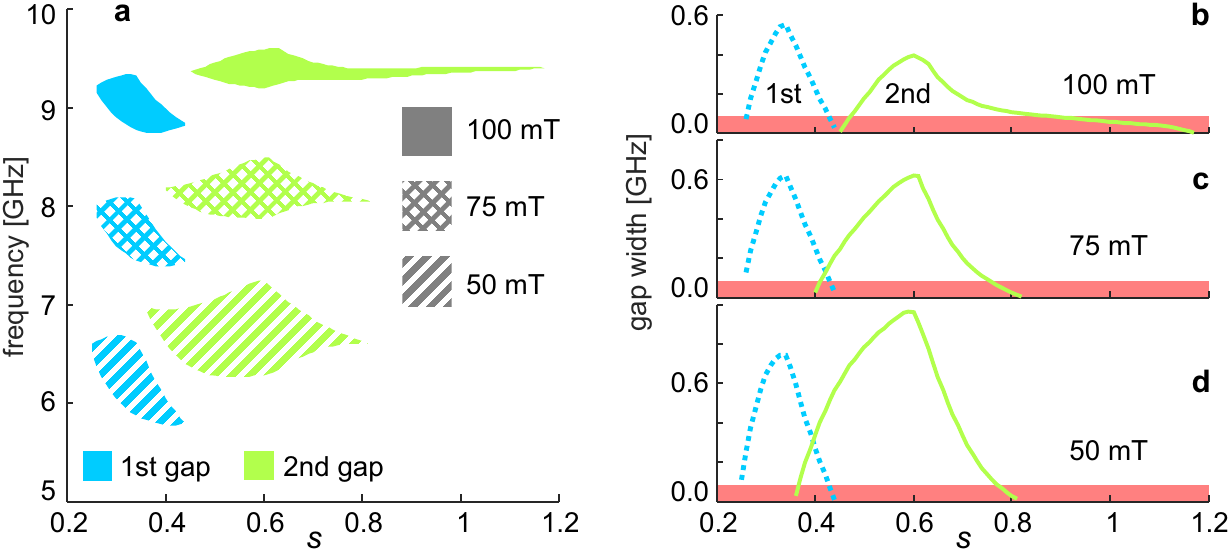}
\caption{(a) Two lowest band gaps vs. squeezing (structure ration, $s$) for Co/Py 2D MC at three magnitudes of the external magnetic field. (b--d) The gap width evolution with $s$ for band gaps presented in (a). Red bars depict the maximal blurring of band gaps caused by the Gilbert damping.}
\label{Fig4}
\end{figure*}

\begin{figure*}
\includegraphics{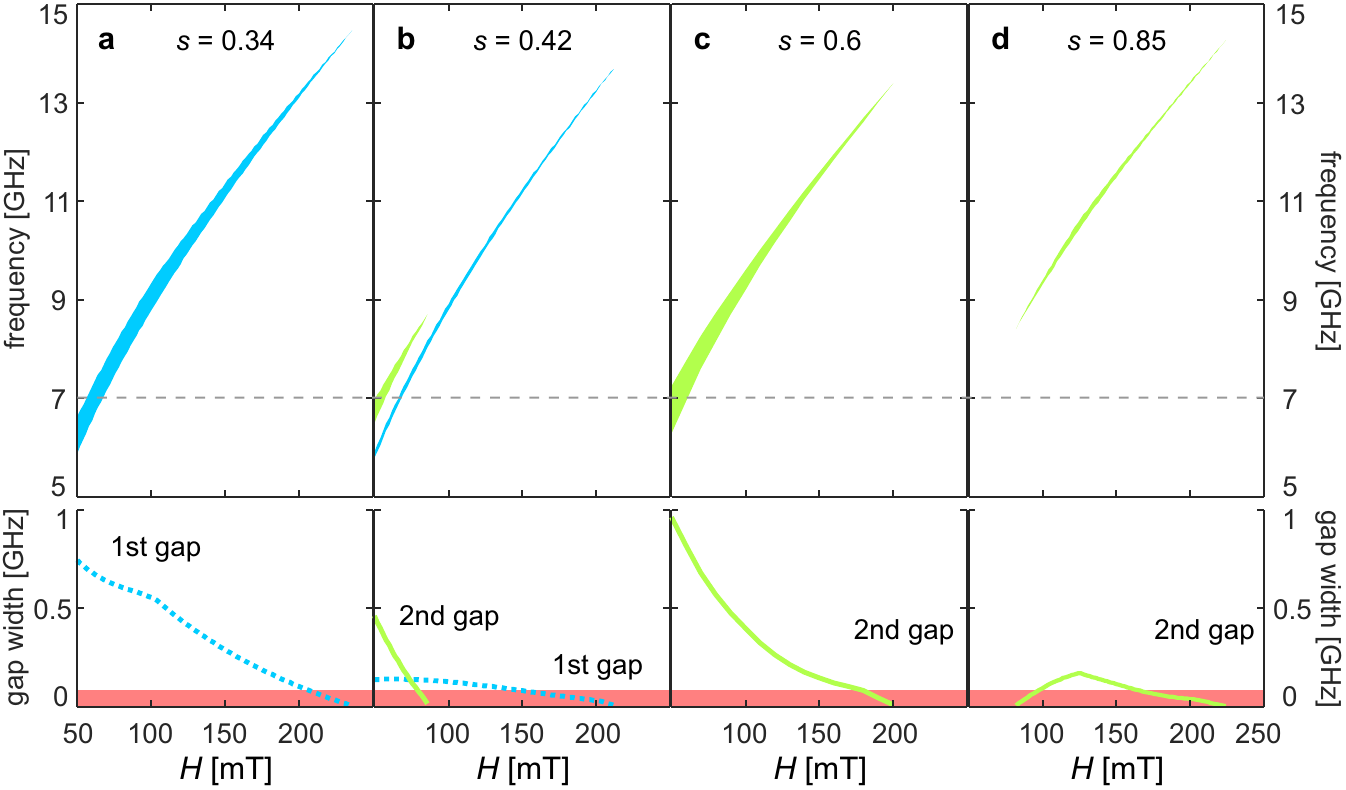}
\caption{Two lowest band gaps vs. the external field magnitude $H$ for four structure ratios: (a) $s = 0.34$, (b) $s = 0.42$, (c) $s = 0.6$, and (d) $s = 0.85$. Below each graph, the dependence of the gap width on $H$ is shown (red bars depict the maximal blurring of band gaps caused by the Gilbert damping). The horizontal dashed line stands for the frequency of 7 GHz (see the text).}
\label{Fig5}
\end{figure*}

In Fig. \ref{Fig4} we illustrate the influence of squeezing on complete magnonic band gaps for three magnitudes of the external magnetic field (50, 75, and 100 mT). Only the two lowest band gaps are presented because in the studied range of an external field other band gaps occur occasionally and their width is restricted to few tens of MHz. Fig. \ref{Fig4}a gives the overall view of the evolution of band gaps with their position and width. Additionally, in Figs. \ref{Fig4}b--d the dependence of the gap width on the squeezing is plotted for both band gaps and all three external field magnitudes (scales are the same in all three graphs). The influence of Gilbert damping is shown as red bars depicting the maximal blurring of band gaps.

For very strong squeezing ($s < 0.25$) there are no band gaps at all. Above this value for all three external field magnitudes the first gap occurs and its width increases with increasing $s$ up to $s = 0.34$ where the maximum of the width is reached. The maximum of the gap width depends on the external field – the highest one is for 50 mT and the lowest for 100 mT. The gap disappears for $s > 0.44$ independently of the external field magnitude thus at the low external field (up to c.a. 200 mT) the range of $s$ for which the 1st gap exists does not depend on the field magnitude. For weaker squeezing ($s > 0.35$) the second gap appears. The structure ratio range of its existence clearly depends on the external field albeit the maximum of its width stands for $s = 0.6$ at all fields. At 100 mT the gap occurs even for the structure stretched instead of squeezed ($s = 1.17$) while at lower fields it disappears for $s > 0.82$. On the other hand, at low fields, there is a range of $s$ for which both band gaps coexist.

These properties allow us to manipulate the pass-bands (bands) and stop-bands (band gaps) of the eventual magnonic filter by {\it in operando} squeezing or stretching of the sample. For instance, if the original structure is based on the $s = 0.4$ MC at $H = 50$ mT the stretching by c.a. 8\% closes the first gap and widens the second gap twice (see Fig. \ref{Fig4}d). One can achieve the opposite effect with 8\% squeezing--the second gap closes and the first one widens twice. Simultaneously the pass-band between these two band gaps is shifted in the frequency scale (see Fig. \ref{Fig4}a).

The possibility of squeezing and stretching of 2D MCs is definitely conceivable at present. There are few types of materials which can be utilized for this purpose, e.g., piezoelectrics \cite{Kholkin_book} and ferroelastics \cite{Szafranski_1993, Mielcarek_2005}. Introducing such materials (or their composites \cite{Allahverdi_2002}) as an element of a substrate, similarly to so called magnon-straintronic systems \cite{Sadovnikov_straintronics}, should lead to `stretchable magnonics'.

The above picture of the dependence of band gaps existence and its width on the squeezing of the MC gives the possibility to design the band gaps behavior also with the external field. The change of the external magnetic field magnitude shifts the spin-wave spectrum in the frequency range and for the high magnetic field the common scenario is that this shift is uniform. The situation is much different in the case of low fields especially when the softening of modes takes place.\cite{Topp_gap, Tacchi_gap, Montoncello_soft} The behavior of the spin-wave spectrum in the low magnetic field exhibits features similar to those caused by the squeezing. We address this problem in the context of the increasing importance of the demagnetizing field with a decreasing external field.

In Fig. \ref{Fig5} the evolution of the two lowest band gaps vs. the external magnetic field is shown for four squeezed structures: $s = 0.34$ (maximal width of the 1st gap), $s = 0.42$ (two band gaps coexisting), $s = 0.6$ (maximal width of the 2nd gap), and $s = 0.85$ (no band gaps below 80 mT). In all cases, both band gaps shift toward high frequencies quickly while $H$ increases, but these structures have a very different sensitivity of the gap width to the external field magnitude. At 50 mT the widest gap appears for $s = 0.6$ (Fig. \ref{Fig5}c). This is the 2nd gap which starts with a width 961 MHz but gets narrow rapidly with increasing $H$; at 100 mT it reaches 397 MHz. For higher magnetic field its narrowing slows down and finally, the gap disappears above 200 mT. Also for $s = 0.34$ (Fig. \ref{Fig5}a) at 50 mT the broad gap occurs: the 1st gap is 744 MHz wide. Its width reduces much more slowly than in the previous case and at 100 mT it becomes 555 MHz. The gap closes at 238 mT. In Fig. \ref{Fig5}b we present an example of the coexistence of both band gaps ($s = 0.42$) at the same frequency. At 50 mT the first gap is as narrow as 137 MHz but its width changes very little up to 100 mT. Here, the gap is 126 MHz wide. It closes above 212 mT. The behavior of the second gap is much different. At 50 mT the gap is 461 MHz wide (3.4 times wider than the 1st gap) but it shrinks rapidly at almost constant speed and closes at 87 mT. In the last example ($s = 0.85$, Fig. \ref{Fig5}d) at 50 mT there is no gap but the 2nd gap appears for higher field. It starts at 82 mT, reaches the maximal width at 124 mT (169 MHz), and vanishes above 224 mT. Again, the narrowing of band gaps due to the Gilbert damping is shown as red bars.

The squeezed MCs makes it possible to manipulate the eventual magnonic filter properties by the external field. Let us consider, for example, a frequency equal to 7 GHz (the horizontal dashed line in Fig. \ref{Fig5}). For $s = 0.34$ at a field below 56 or above 67 mT the frequency in question is in the band so it should propagate with no restrictions while at a field ranging from 56 to 67 mT the frequency fits the band gap and should be strongly suppressed (Fig. \ref{Fig5}a). For $s = 0.42$ this frequency is allowed in the field range 57-66 mT and suppressed in ranges 50-57 mT and 66-70 mT (Fig. \ref{Fig5}b). So, the particular squeezing of the structure allows us to designing the response of the spin-wave spectrum to the external field change.

Gilbert damping has almost no effect on the calculated real parts of the frequencies, but it introduces nonzero imaginary parts, i.e., it causes the broadening of spectral lines. This leads to the blurring of band gaps, which depends on several parameters, such as concentration factor and frequency. In Figs. \ref{Fig4} and \ref{Fig5} we show a maximal uncertainty of band gaps caused by the assumed Gilbert damping. Even for the worst possible scenario, all gaps still exist, although in the narrower range of external field values. This result influences tight band gaps (see, e.g., Figs. \ref{Fig5}b and d) while broad band gaps suffer just a little (Figs. \ref{Fig5}a and c), thus the effect of Gilbert damping does not cancel our findings.

All of the examples presented above are 2D MCs based on a squeezed hexagonal lattice and containing Co elliptical dots immersed in the Py matrix. The usage of the squeezed structure makes it possible to tailor omnidirectional band gaps and their behavior via a demagnetizing field design. Then the external magnetic field can be applied for the reversible control of the band gaps. This makes squeezed MCs a promising candidate for tunable spin-wave filters and transducer design. In fact, all of the considered structures can be fabricated using the current technology. Also, band gaps can be determined using standard experimental techniques such as Brillouin light scattering or transmission measurements with vector network analyzer-ferromagnetic resonance (VNA-FMR).\cite{Krawczyk_Grundler}

\section{Conclusions}

In the paper, we have used the PWM to study theoretically 2D MCs consisting of cobalt rods embedded in the thin-film permalloy matrix and based on the squeezed hexagonal lattice. As demonstrated by our results, omnidirectional band gaps in the spin-wave spectrum can open as a result of both squeezing of the structure and changing of the in-plane external magnetic field magnitude. In both cases, the crucial role is played by the demagnetizing field and the distribution of spin-wave amplitude.

The squeezing along the direction of the external field has a great impact on the magnitude of the demagnetizing field. The negative demagnetizing field in cobalt is stronger if MC is based on a more squeezed structure along the magnetic field direction. This leads to the lowering of the effective field and, as a consequence, to the increasing of the spin-wave concentration in rods. Also, the spatial confinement of the excitation depends on squeezing. The confinement influences spin-wave frequencies with regard to the mode profile which causes the effect of squeezing to be mode-dependent. For propagating spin waves, the profile changes with the direction and the length of the wave vector, which makes the effect $\vec{k}$-dependent and finally changes the bandwidth. This leads to broadening or shrinking of the band in accordance with the mode profile in the FBZ center. At the low external field, the demagnetizing field gains importance so these effects become stronger. The mode- and the $\vec{k}$-dependent softening of spin waves may lead to the reversible opening and closing of different magnonic band gaps without magnetization reversal. 
	
These features make squeezed 2D MCs useful for tailoring the spin-wave spectrum in the context of omnidirectional band gaps existence and the behavior of band gaps with the external field change. As an example we have provided four cases in which 1-2 band gaps can be opened and closed by a change of the external field: a single wide gap or two coexisting band gaps that close with different speed while an external field grows, or one gap that occurs for intermediate fields and disappears if the field is too low or too high. Taking into account Gilbert damping leads to the broadening of bands and thus to a narrowing of band gaps, however our main findings are still valid, even if they are for a narrower range of an external magnetic field. On the other hand, our main results should be independent of the choice of any particular materials as long as the magnetization contrast is ensured to provide a negative demagnetizing field inside dots. All proposed structures should be possible to fabricate with state-of-the-art technology which makes squeezed 2D MCs considerable candidates for designing tunable magnonic devices.


\begin{acknowledgments}
The study has received financial support from the EU's Horizon 2020 research and innovation programme under Marie Sklodowska-Curie GA No. 644348 (MagIC), from the Polish Ministry of Science and Higher Education resources for science in 2017--2019 granted for the realization of an international co-financed project (W28/H2020/2017), and from the National Science Centre of Poland under Grant No. UMO-2016/21/B/ST3/00452.
\end{acknowledgments}




\end{document}